
\input harvmac
\def\form{F. A. Smirnov, {\it Form Factors in Completely Integrable
Models of Quantum Field Theory}, in {\it Advanced Series in Mathematical
Physics} 14, World Scientific, 1992.}
\def\MSS{B. Schroer and T. T. Truong, Nucl. Phys. B144 (1978) 80 \semi
E. C. Marino, B. Schroer, and J. A. Swieca, Nucl. Phys. B200 (1982) 473.}
\def\alt{\tilde{\al}}
\def\rb{\bar{\rho}}
\def\rbt{\tilde{\rb}}
\def\rt{\tilde{\rho}}
\def\emrvac{|\emptyset \rangle }
\def\emlvac{\langle \emptyset |}
\def\dstate{ {}^{\ep_1 \cdots \ep_n} \lva{\th_1 \cdots \th_n} }

\def\mass{{\rm {\hat{m}} }}
\def\ma{\mass}
\def\ha{ {\scriptstyle{\inv{2}} }}
\def\mtoo{{\mathop{\longrightarrow}^{\ma \to 0}}}

\def\betah{{\hat{\beta}}}
\def\hp{\CH_P}
\def\hf{\CH_F}

\def\slh{{\hat{sl(2)}}}
\def\psib{\bar{\psi}}

\def\vphi{\varphi}
\def\bh{\hat{\beta}}

\def\bp{b^+}
\def\bm{b^-}
\def\bb{\bar{b}}
\def\bbp{\bar{b}^+}
\def\bbm{\bar{b}^-}

\def\mn{\frac{ \ma^{|n|+|m|} }{\ma^{|n+m|} } }

\def\slhh{\hat{\hat{sl(2)}}}

\def\lvacp{\langle +\ha \vert}
\def\lvacm{\langle -\ha \vert}

\def\rvacpm{\vert \pm \ha \rangle}
\def\rvacmp{\vert \mp \ha \rangle}
\def\rvacp{\vert + \ha \rangle}
\def\rvacm{\vert - \ha \rangle}
\def\va#1{\vert {#1} \rangle}
\def\lva#1{\langle {#1} \vert}

\def\mat#1#2#3#4{\left(\matrix{#1&#2\cr #3 &#4 \cr}\right)}
\def\col#1#2{\left(\matrix{#1\cr #2\cr}\right)}
\def\dz{\frac{dz}{2\pi i}}
\def\dzb{\frac{d\zb}{2\pi i}}

\def\psipm{\psi_\pm}
\def\psibpm{\psib_\pm }
\def\bpm{b^\pm}
\def\bbpm{\bar{b}^\pm}

\def\hal{{\CH^L_a}}

\def\vhh{\hat{\hat{V}}}

%
%
%

\def\rb{ \right]  }
\def\tilde{\widetilde}
\def\bar{\overline}
\def\hat{\widehat}
\def\*{\star}
\def\[{\left[}
\def\]{\right]}
\def\({\left(}		
\def\){\right)}

%
%
\def\zb{{\bar{z} }}
\def\frac#1#2{{#1 \over #2}}
\def\inv#1{{1 \over #1}}

\def\d{\partial}

\def\rvac{\hbox{$\vert 0\rangle$}}
\def\lvac{\hbox{$\langle 0 \vert $}}
\def\2pi{\hbox{$2\pi i$}}

\def\dsl{\raise.15ex\hbox{/}\kern-.57em\partial}
\def\Dsl{\,\raise.15ex\hbox{/}\mkern-.13.5mu D}
%
%
\def\th{\theta}

\def\al{\alpha}
\def\ep{\epsilon}
\def\la{\lambda}	\def\La{\Lambda}
\def\de{\delta}		
\def\om{\omega}		
	
\def\vphi{\varphi}
%
%
		
		\def\CF{{\cal F}}
	\def\CH{{\cal H}}	
		
		\def\CO{{\cal O}}
		
		\def\CU{{\cal U}}

\def\rvac{\hbox{$\vert 0\rangle$}}
\def\lvac{\hbox{$\langle 0 \vert $}}

\def\2pi{\hbox{$2\pi i$}}

\def\dsl{\raise.15ex\hbox{/}\kern-.57em\partial}
\def\Dsl{\,\raise.15ex\hbox{/}\mkern-.13.5mu D}
%
%
%
\font\numbers=cmss12
\font\upright=cmu10 scaled\magstep1
\def\stroke{\vrule height8pt width0.4pt depth-0.1pt}
\def\topfleck{\vrule height8pt width0.5pt depth-5.9pt}
\def\botfleck{\vrule height2pt width0.5pt depth0.1pt}
\def\Zmath{\vcenter{\hbox{\numbers\rlap{\rlap{Z}\kern
0.8pt\topfleck}\kern
2.2pt
                   \rlap Z\kern 6pt\botfleck\kern 1pt}}}
\def\Qmath{\vcenter{\hbox{\upright\rlap{\rlap{Q}\kern
                   3.8pt\stroke}\phantom{Q}}}}
\def\Nmath{\vcenter{\hbox{\upright\rlap{I}\kern 1.7pt N}}}
\def\Cmath{\vcenter{\hbox{\upright\rlap{\rlap{C}\kern
                   3.8pt\stroke}\phantom{C}}}}
\def\Rmath{\vcenter{\hbox{\upright\rlap{I}\kern 1.7pt R}}}
\def\Z{\ifmmode\Zmath\else$\Zmath$\fi}
\def\Q{\ifmmode\Qmath\else$\Qmath$\fi}
\def\N{\ifmmode\Nmath\else$\Nmath$\fi}
\def\C{\ifmmode\Cmath\else$\Cmath$\fi}
\def\R{\ifmmode\Rmath\else$\Rmath$\fi}
\def\Zmath{Z}


\def\MSS{B. Schroer and T. T. Truong, Nucl. Phys. B144 (1978) 80 \semi
E. C. Marino, B. Schroer, and J. A. Swieca, Nucl. Phys. B200 (1982) 473.}

\def\FrResh{I. B. Frenkel and N. Yu. Reshetikhin, Commun. Math. Phys. 146
(1992) 1. }
\def\VVstar{B. Davies, O. Foda, M. Jimbo, T. Miwa and A. Nakayashiki,
Commun. Math. Phys. 151 (1993) 89;
M. Jimbo, K. Miki, T. Miwa and A. Nakayashiki, {\it Correlation
Functions of the XXZ model for $\Delta < -1$}, Kyoto 1992  preprint,
PRINT-92-0101,
hep-th/9205055.}

\def\BLnlc{D. Bernard and A. LeClair, Commun. Math. Phys. 142 (1991) 99;
Phys. Lett. B247 (1990) 309.}
\def\form{F. A. Smirnov, {\it Form Factors in Completely Integrable
Models of Quantum Field Theory}, in {\it Advanced Series in Mathematical
Physics} 14, World Scientific, 1992.}

\def\Frenkeli{I. B. Frenkel and N. Jing,
Proc. Natl. Acad. Sci. USA 85 (1988) 9373.}
%
%
%

%
%
%

%

%

%

%
%
%

%

\Title{CLNS 95/1318}
{\vbox{\centerline{Form Factors from Vertex Operators and    }
\centerline{ Correlation Functions at $q=1$} }}

\bigskip
\bigskip

\centerline{Andr\'e LeClair}
\medskip\centerline{Newman Laboratory}
\centerline{Cornell University}
\centerline{Ithaca, NY  14853}
\bigskip
\centerline{Proceedings of SMQFT Conference, USC, Los Angeles, May 1994}

\bigskip

\vskip .3in

We present a new application of affine Lie algebras to
massive quantum field theory in 2 dimensions, by investigating
the $q\to 1$ limit of the q-deformed affine $\hat{sl(2)}$  symmetry of
the sine-Gordon theory, this limit occurring at the free fermion point.
We describe how radial quantization leads to a quasi-chiral factorization of
the space of fields.
The
conserved charges which generate the
affine Lie algebra split into two independent affine algebras on
this factorized space, each with level 1, in the anti-periodic sector.
The space of fields in the anti-periodic sector can be organized
using level-$1$ highest weight representations, if one supplements
the $\slh$ algebra with the usual local integrals of motion.
Using the integrals of motion, a momentum
space bosonization involving vertex operators is formulated.
This leads to a new way of computing form-factors, as
vacuum expectation values  in momentum space.
The problem of non-trivial correlation functions in this model is
also discussed;  in particular it is shown how
space-time translational anomalies which arise in radial quantization
can be used to compute the short distance expansion of some simple
correlation functions.

\Date{1/95}
%
%
%
%
%
\noblackbox
\def\ot{\otimes}
\def\zb{{\bar{z}}}

\def\zbar{{\bar{z}}}

%
%
%
%
%
%
%
%
%
%

\newsec{Introduction}

It is well known that massive integrable quantum field theories
in two dimensions are characterized by an infinite number of local
integrals of motion $P_n$ which all commute.  Under Lorentz transformations
these integrals of motion are characterized by their integer Lorentz
spin
$$
\[ L , P_n \]  = n ~ P_n  $$
where $L$ is the Lorentz boost operator.  For example, in the
sine-Gordon theory, there exist $P_n$'s  for all odd integer $n$.
The local integrals of motion are important for establishing
integrability and consequently the factorizability of the S-matrix.
They  also play a central role in the quantum inverse scattering method.

Since the local integrals of motion satisfy an absolutely uninteresting
infinite abelian algebra, they provide only limited information
on the S-matrix, and have not provided any useful Ward identities
for correlation functions.  More recently, attention has turned to
the existence of infinite non-abelian symmetries.  Generally, these
correspond to $q$-deformed affine Lie algebras and Yangians.
The extent to which these quantum affine symmetries characterize
the  dynamical properties of the models remains an open question.
The main dynamical properties one is interested in are the S-matrix,
the form factors and the correlation functions.  The fact that the
S-matrix is almost completely characterized by its quantum affine
symmetry suggests it is worthwhile to understand the consequences of this
symmetry for the form factors and correlation functions.

We will consider the sine-Gordon theory, characterized by the
lagrangian
\eqn\eIi{
S = \inv{4\pi} \int d^2 z \(  \d_z \phi \d_\zb \phi
                 + 4\la  \cos ( \betah \phi ) \) . }
In \ref\rbl{\BLnlc}\ref\rfl{G. Felder and A. LeClair,
Int. Journ. Mod. Phys. A7 Suppl. 1A (1992) 239.},
explicit conserved currents were constructed for the finite number
of simple roots
of the
 $q-\hat{sl(2)}$ affine Lie algebra, where
$q = \exp ( -2\pi i / \betah^2 )$.
The main features of these currents are that they are non-local
and possess fractional Lorentz spin $\pm (2/\bh^2 -1 )$.
All of the Hopf algebraic properties of the quantum affine symmetry
are a consequence of this non-locality, in particular, $q^2$
is a braiding phase.   The non-locality of these conserved charges
is precisely what makes it difficult to use them to derive Ward
identities for example.   To simplify the problem, consider
the `reflectionless' points $\bh^2 = 2/(N+1)$.  Here, $q^2 = 1$,
and the Lorentz spin of the charges is $N$.
One is thus led to suspect the existence of an undeformed affine
Lie algebra symmetry $\hat{sl(2)}$ at all the reflectionless points,
however this does not follow from results in \rbl\ due to the
delicacy of the $q^2 \to 1 $ limit.
In this talk we will consider only the case of $N=1$, which occurs
at the free fermion point.
Indeed, one can construct explicitly all of the generators for the
$\hat{sl(2)}$ symmetry in this case.
The aim of this work is to  understand
the extent to which the affine symmetry characterizes the dynamical
properties of the model, and to develop some new structures that may
be amenable to $q$-deformation.  The deformation of the results presented
here away from the free fermion point still remains a difficult open
problem.
Some aspects of the quantum affine symmetry at general reflectionless
points is studied in \ref\rln{A. LeClair and D. Nemeschansky,
{\it Affine Lie Algebra Symmetry of Sine-Gordon Theory at
Reflectionless Points}, to appear.  }

Most of the work presented here is based on the papers\ref\rlec{A.
LeClair, NPB 415 (1994) 734.}\ref\rcostas{C. Efthimiou and A. LeClair,
{\it Particle-Field Duality and Form Factors from Vertex Operators},
to appear in Commun. Math. Phys., hep-th/9312121.}.  Only the last
section contains previously unpublished work.

\newsec{$\hat{sl(2)}$ Symmetry of Massive Dirac Fermions}

In this section we describe the conserved currents for the
affine $\hat{sl(2)}$ symmetry at the free fermion point.
It is well known that at $\bh = 1$, the sine-Gordon theory
is equivalent to a massive free field theory of charged fermions.
Introducing the Dirac spinors
$ \Psi_\pm = \left(\matrix{\psib_\pm \cr \psi_\pm \cr}\right)$
of $U(1)$ charge $\pm 1$,  the equations of motion are
\eqn\eIIxi{
\d_z \psib_\pm = i \ma \psi_\pm ,~~~~~
\d_\zb \psi_\pm = -i \ma \psib_\pm . }
We have continued to Euclidean space, and $z,\zb$ are the
usual Euclidean light-cone coordinates:
$
z = (t + ix)/2 , ~~~~~\zb = (t-ix)/2 $.

Generally, the conserved quantities $Q$ follow from conserved currents
$J_\mu$:
\eqn\eIIiv{
\d_\zb J_z  + \d_z J_\zb = 0;   ~~~~~~~~
Q = \int \dz ~  J_z ~ - ~ \int \dzb ~ J_\zb . }
Define the inner product of two spinors
$A = \col{\bar{a}}{a}$, $B = \col{\bar{b}}{b}$ as
\eqn\eVii{
(A,B) = \inv{4\pi} \( \int  dz
{}~ a \>  b ~+~ \int d\zbar  \bar{a} \> \bar{b} \) . }
Using the equations of motion
\eIIiv, one can verify that the following charges are conserved:
\eqn\eVv{\eqalign{
Q^\pm_{-n} = \inv{2}  \( \Psi_\pm , \d_z^n \Psi_\pm \)& , ~~~~~
Q^\pm_{n} = \inv{2} \( \Psi_\pm , \d_\zb^n \Psi_\pm \) ~~~~~ n ~{\rm odd}
\cr
 \al_{-n} = : \( \Psi_+ , \d_z^n \Psi_- \) : & , ~~~~~
 \al_{n} = : \( \Psi_+ , \d_\zb^n \Psi_- \) :
 \cr }}
where $n\geq 0$.

Define
$$P_n = \al_n , ~~~n ~{\rm odd}; ~~~~~~T_n = \al_n , ~~~n~{\rm even}.$$
Then one can show that the charges obey the following commutation
relations:
\eqna\eIIxxiii
$$\eqalignno{
\[ P_n , P_m \] &= 0  &\eIIxxiii {a} \cr
\[ P_n , T_m \] &= \[ P_n , Q^\pm_m \] = 0 &\eIIxxiii {b} \cr
\[ T_n , T_m \] &= 0 &\eIIxxiii {c} \cr
\[ T_n , Q^\pm_m \] &= \pm 2 ~ \mn ~ Q^\pm_{n+m}  &\eIIxxiii {d} \cr
\[ Q^+_n , Q^-_m \] &= \mn ~ T_{n+m} &\eIIxxiii {e} . \cr } $$
The conserved charges have integer Lorentz spin:
\eqn\eIIxxv{
\[ L , \al_n \] = -n \> \al_n , ~~~~~\[ L , Q^\pm_n \] = -n \> Q^\pm_n . }
The $P_n$ are the usual local integrals of motion at odd Lorentz spin,
where the translation operators are $P_z = P_{-1}, ~~P_{\zbar} = P_1$.
The remaining generators $T_n , ~ Q^\pm _n $ satisfy the {\it twisted}
$\hat{sl(2)}$ algebra at level $0$, and furthermore commute with the
$P_n$.  The twist is to be expected here since in the free fermion
theory only the U(1) charge is Lorentz spinless.  The twist breaks
the sl(2) subalgebra of zero modes that is familiar in current algebra
to U(1).

It is interesting to consider the conformal limit of the above
charges.  As $\ma \to 0$,  the charges $Q_{-n}$ for $n\geq 0$
become purely left-moving whereas the $Q_n$ become right-moving.
Thus, in the conformal limit one recovers two decoupled Borel
subalgebras of $\hat{sl(2)}$.

The lowest conserved charges $Q^\pm_1 , ~ Q^\pm_{-1} $  can be
derived from the results in \rbl\ using standard bosonization
formulas, in fact this is how we originally found them.
See also \ref\raj{R. K. Kaul and R. Rajaraman, Int. J. Mod. Phys.
A8 (1993) 1815}\ref\abda{E. Abdalla, M. C. B. Abdalla, G. Sotkov
and M. Stanishkov, {\it Off-critical Current Algebras}, Univ.
Sao Paulo Preprint, IFUSP-preprint-1027, Jan. 1993.}

\newsec{Representation Theory}

In this section we consider the representations of $\hat{sl(2)}$
that are realized in the model.  There are two vector spaces of
interest: the space of particles and the space of fields.
The space of particles diagonalizes the translation operators,
which on one particle states have the eigenvalues
$P_z = \ma e^\theta , ~~P_\zbar = \ma e^{-\theta}$.  The space of
particles is then:
$$
\CH_P = \{ \oplus | \theta_1 \cdots \th_n \rangle \} . $$
On the doublet of one-particle states, the conserved charges
have the following loop algebra (level 0) representation:
\eqn\eIIxxxii{
\al_n = \ma^{|n|} u^{-n} \mat{1}{0}{0}{ (-1)^{n+1} } , ~~~
Q^+_n = \ma^{|n|} u^{-n} \mat{0}{1}{0}{0} , ~~~
Q^-_n = \ma^{|n|} u^{-n} \mat{0}{0}{1}{0} . }
The trivial comultiplication induces a representation of
$\hat{sl(2)}$ on the tensor product space $\CH_P$.
Thus $\CH_P$ is a direct sum of finite dimensional level 0 representations
of $\hat{sl(2)}$.

Consider now the space of fields.  Let $\Phi_i (z, \zbar )$ denote
a complete basis in the space of fields and let
$|\Phi_i > = \Phi_i (0) |0> $.  The space of fields is then defined as
\eqn\fields{
\CH_F = \{ \oplus_i  |\Phi_i >  \}  .}
The space of fields of course does not diagonalize the translation
operators $P_z , P_\zbar$, but rather diagonalizes the Lorentz boost
operator $L$.   The representions of $\hat{sl(2)}$ on the space of
fields are much more interesting than on the space of particles.
An explicit construction of the space of fields can be obtained
via radial quantization, since it is this quantization scheme that
yields states that diagonalize $L$.  In this way one can show
that there exists a quasi-chiral splitting in the space of fields:
$$\CH_F = \CH_F^L \otimes \CH_F^R . $$
In the conformal limit, $\CH_F^{L,R}$ are the left and right moving
chiral sectors;  however we emphasize that radial quantization
provides a meaning to the quasi-chiral splitting in the massive
theory.  Furthermore, as I will describe, the conserved charges
split into  quasi left and right pieces, and the spaces
$\CH_F^{L,R}$ turn out to be infinite dimensional level 1 representations
of $\hat{sl(2)}$.

We now describe the basic features of radial quantization that allows
one to establish the above statements.  In radial quantization,
equal `time' surfaces are taken to be circles surrounding the origin
in the space-time plane.  For the free fermion, one only needs
to solve the Dirac equation in polar coordinates to obtain the
expansions of the fields appropriate to radial quantization.
One finds
\eqn\eIIIxvii{
\col{\psibpm}{\psipm} = \sum_{\om }
\bpm_\om ~ \Psi^{(a)}_{-\om - 1/2} ~+~
\bbpm_\om ~ \bar{\Psi}^{(a)}_{-\om -1/2} . }
The labels $a,p$ represent anti-periodic versus periodic sectors.
In the anti-periodic sector, $\om \in \Zmath$, whereas in the
periodic sector $\om \in \Zmath + 1/2$.
We will mainly be concerned with the anti-periodic sector,
where one has:
\eqn\eIIIxviii{\eqalign{
\Psi^{(a)}_{-\om -1/2} &=
\Gamma (\ha - \om ) ~ \ma^{\om + 1/2} ~
\col{i \> e^{i(\ha - \om )\vphi} ~ I_{\ha - \om} (\ma r) }
{e^{-i(\om + \ha )\vphi} ~ I_{-\om -\ha} (\ma r)}  \cr
\bar{\Psi}^{(a)}_{-\om - 1/2}
&=
\Gamma (\ha - \om ) ~ \ma^{\om + 1/2} ~
\col{ e^{i(\ha + \om )\vphi} ~ I_{-\ha - \om} (\ma r) }
{-i \> e^{-i(\ha - \om )\vphi} ~ I_{\ha - \om } (\ma r)}  .\cr}}
Similar formulas apply to the periodic sector.

Upon quantization, one finds
\eqn\eIIIxxiv{
\{ \bp_\om , \bm_{\om'} \} = \de_{\om , -\om'}
, ~~~~~\{ \bbp_\om , \bbm_{\om'} \} = \de_{\om , -\om'}
, ~~~~~\{ b_\om , \bb_{\om '} \} = 0. }
Since the $b$ operators commute with the $\bar{b}$ operators,
the fermionic Fock space built out of the the fermion modes
factorizes as explained above.

The fermionic Fock spaces constructed using the operators
$b_\om , ~ \bar{b}_\om$  can be identified with the space of fields.
For example, in the periodic sector one finds
\eqn\eIIIxxxi{
\d_z^n \psipm (0) \rvac = n! ~ \bpm_{-n-\ha} \rvac , ~~~~~
\d_\zb^n \psibpm (0) \rvac = n! ~  \bbpm_{-n-\ha} \rvac . }
\eqn\eIIIxxxiii{
J_z (0) \rvac = \bp_{-\ha} \> \bm_{-\ha} \> \rvac ,
{}~~~~~J_\zb (0) \rvac = \bbp_{-\ha} \> \bbm_{-\ha} \> \rvac , }
where $J_\mu$ is the U(1) current.   The above formulas are
identical to the conformal limit, however we emphasize that
they are valid in the massive theory:  they are a consequence of
the $r\to 0$ properties of the Bessel functions in the expansion
of the fields.

In the anti-periodic sector, due to the existence of the zero modes
$b_0$, $\bar{b}_0$, there are doubly degenerate Ramond vacuua for
each sector  $|\pm 1/2 >_L$ and $|\pm 1/2 >_R$.
One has the following identification with fields in the sine-Gordon
theory:
\eqn\eIIIxvi{
e^{\pm i \phi (0)/2 } \> \rvac ~=~
(c\ma )^{1/4} ~ \( \rvacpm_L \ot \rvacmp_R \)
\equiv (c\ma )^{1/4} ~ \va{\pm \ha}
 , }
where $c$ is some constant.
The quasi-chiral spaces of fields described above have the following
explicit realization:
\eqn\eIIIxxxvb{\eqalign{
\CH^L_{a_{\pm}}
&= \left\{ \bm_{-n_1} \bm_{-n_2} \cdots \bp_{-n_1 '} \bp_{-n_2 '}
\cdots \rvacpm_L \right\} \cr
\CH^R_{a_{\pm}}
&= \left\{ \bbm_{-n_1} \bbm_{-n_2} \cdots \bbp_{-n_1 '} \bbp_{-n_2 '}
\cdots \rvacpm_R \right\} , \cr}}
for $n , n' \geq 1$.

In the radial quantization the conserved charges split into quasi
left and right pieces:
\eqn\eVvii{\eqalign{
Q^\pm_n ~&=~ Q^{\pm , L}_n ~+~ Q^{\pm ,R}_{-n}  \cr
\al_n ~&=~ \al^L_n ~+~ \al^R_{-n}  . \cr }}
The $Q^L$ ($Q^R$) pieces involve only the $b$ ($\bb$) operators.
For instance,
\eqn\eqrad{
Q^{\pm , L}_n = \ma^{|n| +n} \sum_{\om \in \Zmath}
\frac{\Gamma(\ha + \om - n )}{\Gamma (\ha + \om ) } ~
b^\pm_{n-\om} b^\pm_\om . }
With the above expression for the conserved charges in terms of the
fermion modes, one can study the representation of the charges on
the space of fields.  One finds that the split charges now satisfy
a level 1 affine Lie algebra:
\eqn\eVxvii{\eqalign{
\[ P^L_n , P^L_m \] &=  n ~ \ma^{2|n|} ~ \de_{n, -m}  \cr
\[ P^L_n , T^L_m \] &= \[ P^L_n , Q^{\pm,L}_m \] = 0  \cr
\[ T^L_n , T^L_m \] &= n~ \ma^{2|n|} ~ \de_{n, -m}  \cr
\[ T^L_n , Q^{\pm ,L}_m \] &= \pm 2 ~ \mn ~ Q^{\pm, L}_{n+m}   \cr
\[ Q^{+,L}_n , Q^{-,L}_m \] &= \mn ~ T^L_{n+m} ~+~
\frac{n}{2} ~ \ma^{2|n|} ~ \de_{n,-m}  , \cr}}
and similarly for the right charges.  The non-zero levels cancel
in the sum $Q^L + Q^R$ so that the complete charges continue to
satisfy a level 0 algebra.

Let us denote the $P_n$ extension of the algebra $\slh$ defined in
\eVxvii\ as $\slhh$.  As we have seen, the symmetry algebra factorizes
into $\slhh_L \ot \slhh_R$.  The algebra $\slhh$ is
a complete spectrum generating algebra for the anti-periodic sector, namely,
the complete spectrum of quasi-chirally factorized fields can be
obtained from infinite highest weight representations of $\slhh$.
To show this define modules
\eqn\eVxxvi{\eqalign{
\vhh^{{}_L}_{\La_0} &\equiv \left\{ Q^{\pm,L}_{-n_1} ~Q^{\pm , L}_{-n_2} \cdots
T^L_{-n_1 '} T^L_{-n_2 '} \cdots P^L_{-n_1 ''} P^L_{-n_2 ''} \cdots
\rvacm_L \right\} \cr
\vhh^{{}_L}_{\La_\ha}
&\equiv \left\{ Q^{\pm,L}_{-n_1} ~Q^{\pm , L}_{-n_2} \cdots
T^L_{-n_1 '} T^L_{-n_2 '} \cdots P^L_{-n_1 ''} P^L_{-n_2 ''} \cdots
\rvacp_L \right\} , \cr}}
for $n,n' , n'' \geq 1$.
Then, using characters one can show that
\eqn\eident{
\hal = \vhh^{{}_L}_{\Lambda_0} \oplus \vhh^{{}_L}_{\Lambda_\ha} . }

\newsec{Particle-Field Duality and Form-Factors from Vertex Operators}

\def\rb{\bar{\rho}}
\def\rbt{\tilde{\rb}}
\def\rt{\tilde{\rho}}
\def\emrvac{|\emptyset \rangle }
\def\emlvac{\langle \emptyset |}

Form factors are matrix elements of fields in the space of states
$\hp$.  The basic form factors
$$\dstate \Phi (0) |0>  , $$
from which the more general matrix
elements may be obtained by  crossing symmetry, are inner products
of states in $\hf$ with states in $\hp^*$.  (Here,
the indices $\ep_i$ are isotopic.)  The completeness relation
in the space of particles,
\eqn\eIIiv{
1 = \sum_{\vec{\th}} \va{\overrightarrow{\th}} \lva{\overleftarrow{\th}}
=
\sum_{n=0}^\infty \inv{n!} \sum_{ \{ \ep_i \} } \int d\th_1 \cdots
d\th_n ~ |\th_1 ,\cdots , \th_n \rangle_{\ep_1 \cdots \ep_n}
{}^{\ep_n \cdots \ep_1 } \langle \th_n , \cdots , \th_1 | . }
allows us to map states in $\hf$ to states in $\hp$, i.e. to
view $\va{\Phi} \in \hp$:
\eqn\eIIvi{
\va{\Phi_i} =
 \sum_{\vec{\th}} \va{\overrightarrow{\th}}
 \lva{\overleftarrow{\th}} \Phi_i \rangle .}
The intuitive simplicity of the space $\hp$ is responsible for this
conventional way of thinking about form factors.

We give now a dual description of the same form factors.  Let us
suppose that one can define a dual to the space of fields $\hf^*$
with inner product and completeness relation:
\eqn\eIIvii{\eqalign{
\lva{\Phi^i} \Phi_j \rangle &= \delta^i_j  \cr
1 &= \sum_i \va{\Phi_i} \lva{\Phi^i }  . \cr }}
Then one can map a state $\va{\vec{\th}} \in \hp$ into
$\hf$.  The dual statement is
\eqn\eIIviii{
\dstate = \sum_{\Phi_i \in \CF}
\dstate \Phi_i \rangle \lva{\Phi^i } ~~~~\in \hf^* . }

In order to work the above simple remarks into an efficient means
of computing form factors, one must work explicitly with the space
$\hf$.
Radial quantization provides the necessary structures \eIIvii.
In order to use these ideas to compute form factors, we need to
introduce the notion of vertex operators.  The formula
\eIIviii\ implies that one can map states in $\hp^*$ to states
in $\hf^*$.  We call this map the `particle-field map'.
We construct this map explicitly by defining vertex
operators  $V^\ep (\th )$ as follows:
\eqn\eIIix{
\dstate = \lva{\Omega} ~ V^{\ep_1} (\th_1 ) \cdots
V^{\ep_n} (\th_n ) ~~ \in \hf^* }
where $\lva{\Omega}$ is a fixed `vacuum' state, which
will be characterized completely below. The
vertex operators are distinguished from the Faddeev-Zamolodchikov
operators $Z(\th )$ since they act on completely different spaces.
However the basic algebraic relations satisfied by the $Z$ operators
continue to be satisfied by the $V$ operators.
The vertex operators $V^{\ep} (\th )$ operate in the space $\hf$:
\eqn\eIIx{
V^{\ep} (\th ) : ~~~~\hf \to \hf . }
In the sequel we will describe how to construct these vertex
operators explicitly using radial quantization.  Once the
vertex operators are constructed, the form factors
$\dstate \Phi_i \rangle$ are computed directly in the
discrete space $\hf$ using \eIIix.

The vertex operators can be  constructed explicitly using the
radial modes introduced in the last section.
Let
$$ u = e^\theta ,$$
and define
\eqn\eIIIxiv{\eqalign{
\bpm (u) &= \pm i \sum_{\om \in \Zmath } \Gamma (\ha - \om )
\> \ma^\om ~ b^\pm_\om ~ u^\om   \cr
\bbpm (u) &= \pm  \sum_{\om \in \Zmath } \Gamma (\ha - \om )
\> \ma^\om ~ \bb^\pm_\om ~ u^{-\om}  .    \cr }}
Then the vertex operators are
\eqn\eIVxii{\eqalign{
\lva{\Omega}  &= \lva{\ha} + \lva{-\ha}  \cr
V^\ep (\th )  &=  \inv{\sqrt{2\pi^2 i } }
\( b^\ep (e^{-i\pi} u )  + \bb^\ep ( e^{-i\pi} u ) \) . \cr}}
The normalization of the vertex operators was chosen to satisfy the
residue property
\eqn\eIVxiii{
V^+ (\th ) V^- (\th + \beta + i\pi ) ~\sim ~  \inv{i\pi \> \beta} , }
which leads to the proper residue axiom for the multiparticle
form factors.  From \eIIIxvi, one sees that the choice \eIVxii\ for
$\lva{\Omega}$ is equivalent to  the following vacuum expectation values:
\eqn\eIVxiv{
\lvac ~ e^{\pm i \phi (0)/2} ~ \rvac = (c\ma )^{1/4} }

One can use the above construction to compute the form factors
of the fields $\exp (\pm i \phi /2 )$.  For these fields,
all of the form factors with a $U(1)$ neutral combination of an
even number of particles is non-zero.  The result is
\eqn\effii{\eqalign{
& {}^{+++...---...} \lva{\th_1 , \th_2 , \cdots , \th_{2n} }
{}~ e^{\pm i \phi (0) /2 } \rvac \cr
&~~~~=
(c\ma )^{1/4}~    \langle \mp \ha \vert
V^+ (u_1 ) \cdots V^+ (u_n ) V^- (u_{n+1} ) \cdots V^- (u_{2n} )
\vert \pm \ha \rangle \cr
&~~~~= (c\ma )^{1/4} \frac{ (\pm 1)^n }{( i\pi )^n } (-1)^{n(n-1)/2}
\sqrt{u_1 \cdots u_{2n} }
\( \prod_{i=1}^n \( \frac{u_{i+n}}{u_i} \)^{\pm 1/2} \)
\( \prod_{i<j \leq n} (u_i - u_j ) \)  \cr
& ~~~~~~~~\times \( \prod_{n+1 \leq i < j } (u_i - u_j ) \)
\( \prod_{r=1}^n \prod_{s=n+1}^{2n} \inv{u_r + u_s } \)   . \cr } }
The above computation can be done using the Wick theorem with the
2-point functions
\eqn\eIVvii{\eqalign{
{}_L \lvacm ~ b^+ (u) \> b^- (u' ) ~ \rvacp_L
= {}_R \lvacp ~ \bb^+ (u) \> \bb^- (u' ) ~ \rvacm_R  &= \pi
\frac{u'}{u+u'} \cr
{}_L \lvacp ~ b^+ (u) \> b^- (u' ) ~ \rvacm_L
= {}_R \lvacm ~ \bb^+ (u) \> \bb^- (u' ) ~ \rvacp_R  &= - \pi
\frac{u}{u+u'} . \cr}}
However the computation is more easily
done using the bosonization techniques of
the next section.
After some algebraic manipulation,
one can see that these expressions agree with the known results,
though they were
originally computed using rather different
methods\ref\rmss{\MSS}\ref\rform{\form}.

\newsec{Bosonization in Momentum Space}

In the massive theory one can use the constants of
motion to formulate an exact bosonization.
In the anti-periodic sector, since $\al^{L,R}_n$ satisfy two separate
Heisenberg algebras, they can be used to construct a bosonization.
Define
$$\rho_n = \ma^{-|n|} \al^{L}_n , ~~~~~~\rb_n = \ma^{-|n|} \al^R_n  ,$$
and define the momentum space fields (recall $u= e^\th$):
\eqn\eVIx{\eqalign{
-i \rho (u) &=    \sum_{n\neq 0}  ~ \rho_n  ~ \frac{u^n}{n}
  + \rho_0 \log (u) - \rt_0  \cr
-i \rb  (u) &=    \sum_{n\neq 0}  ~\rb_n  ~ \frac{u^{-n}}{n}
   - \rb_0 \log (u) - \rbt_0  , \cr }}
where one  has $[ \rho_0 , \tilde{\rho}_0 ] = [\rb_0 , \rbt_0 ] = 1$.
We further define an auxiliary vacuum satisfying
\eqn\avac{
\al_n^L \emrvac = \al_n^R \emrvac = 0 , ~~n\geq 0; ~~~~~
\alt^L_0 \emrvac ~, ~~\alt^R_0 \emrvac \neq 0 . }
This vacuum $\emrvac$ is not to be confused with the physical
vacuum $\rvac$ which resides in the periodic sector.  One has
the following  expectation values:
\eqn\eVIxiv{\eqalign{
 \emlvac ~ \rho (u) ~ \rho (u') ~\emrvac
&= -\log ( 1/u - 1/u' ) \cr
 \emlvac ~ \rb (u) ~ \rb (u') ~\emrvac  &= -\log (u - u' ) \cr}}
\medskip
\eqn\eVIxv{\eqalign{
 \emlvac \prod_i ~ e^{i\al_i \rho (u_i )} ~ \emrvac
&= \prod_{i< j} \( 1/u_i - 1/u_j \)^{\al_i \al_j }  \cr
 \emlvac \prod_i ~ e^{i\al_i \rb (u_i )} ~ \emrvac
&= \prod_{i< j} \( u_i - u_j \)^{\al_i \al_j }  . \cr}}
The bosonized expressions for the operators
$b^\pm (u) , \bb^\pm (u)$ and the states $\va{\pm \ha}$
follow from the basic commutation relations
\eqn\eVxvi{\eqalign{
\[ \al_n^L , b^\pm (u) \] &= (\pm 1)^{n+1} \ma^{|n|} u^{-n} ~ b^\pm (u) \cr
\[ \al_n^R , \bb^\pm (u) \] &= (\pm 1)^{n+1}
\ma^{|n|} u^{n} ~ \bb^\pm (u) \cr }}
and the 2-point functions \eIVvii.  The commutation relations
\eVxvi\ are fundamental in the sense that they describe how the
conserved charges are represented on asymptotic multiparticle
states.  One finds
\eqn\eVvii{
\sqrt{ \frac{\pm u}{\pi} }
{}~ b^\pm (u)   = : e^{\pm i \rho (\pm u)} :
, ~~~~~~
\frac{\pm 1}{\sqrt{\pm \pi u }}
{}~  \bb^\pm (u)  = : e^{\pm i \rb (\pm u )} :
}
where $-u = e^{-i\pi} u $, and
\eqn\eVIxvi{\eqalign{
\va{\pm \ha}_L ~ &= ~ : e^{\pm i \rho (\infty )/2 } :  \emrvac_L ,
{}~~~~~~~~~~~~~
\va{\pm \ha}_R ~ = ~: e^{\pm i \rb (0) /2 }:  \emrvac_R \cr
{}_L
\lva{\pm \ha} &= \lim_{u\to 0} ~ u^{-1/4} ~ \emlvac :
e^{\pm i \rho (u)/2 } :
, ~~~~~
{}_R
\lva{\pm\ha} = \lim_{u\to \infty} ~ u^{1/4} ~ \emlvac
: e^{\pm i \rb (u) /2 } :
.\cr}}
One can easily check that this construction reproduces the
form factors \effii.

\newsec{Differential Equations for Correlation Functions}

The problem of correlation functions in massive integrable models
is notoriously difficult and generally unsolved.
Even here in the free fermion theory correlators of the
fields $\exp (i \al \phi )$ for general $\al$ are rather complicated.
For $\al = \pm 1/2$ these correlation functions are related to
squares of Ising correlators\ref\riz{J. B. Zuber and C. Itzykson,
Phys. Rev. D15 (1977) 2875.}\rmss.   From the celebrated Ising
results of Wu et. al. \ref\rwu{T. T. Wu, B. M. McCoy, C. A. Tracy
and E. Barouch, Phys. Rev. B13 (1976) 316.}  one can obtain
differential equations for these special sine-Gordon correlators.

In \ref\rfred{D. Bernard and A. LeClair, Nucl. Phys. B426 (1994)
534.}\ differential equations for derived for arbitrary $\al$
in the following way.  Consider the two point function
\eqn\etwopt{
<0| e^{i\al \phi (z, \zb ) } ~ e^{i\al' \phi (0) } |0> .}
Inserting the resolution of the identity
\eIIiv\ between the fields and using the form-factors,
one finds that this correlator
can be written as a Fredholm determinant:
\eqn\efred{
<0| e^{i\al \phi (z, \zb ) } ~ e^{i\al' \phi (0) } |0>
= Det (1 + \bf {K} ) , }
where $\bf{K}$ is a 2 by 2 matrix of integral operators:
\eqn\kernel{
\bf {K}  (u,v) = \left(
\matrix{ 0 & \frac{e(u) \hat{e} (v) }{u+v} \cr
\frac{\hat{e} (u) e(v)}{u+v} & 0 \cr } \right)
}
where
\eqn\kerb{\eqalign{
e(u) &= \( \frac{\sin (\pi \al )}{\pi} \)^{1/2}
u^{(1+\al' - \al )/2 }
\exp \[ -\ha m (zu + \zb u^{-1} ) \]  \cr
\hat{e}(u) &= \( \frac{\sin (\pi \al' )}{\pi} \)^{1/2}
u^{(1+\al - \al' )/2 }
\exp \[ -\ha m (zu + \zb u^{-1} ) \]  \cr
. }}

Using a generalization of the techniques developed in \ref\kora{A. R. Its,
A. G. Izergin, V. E. Korepin and N. A. Slavnov, Int. J. Mod. Phys. B4
(1990) 1003.}\ref\book{V. E. Korepin, N. M. Bogoliubov and A. G.
Izergin, {\it Quantum Inverse Scattering Method and Correlation
Functions}, Cambridge University Press, Cambridge, 1993.} one can
show that the above two point functions are parameterized in terms
of a solution $\eta$  of the sinh-Gordon equation\rfred.
Let
$$ \Sigma (z, \zb ) = \log
<0| e^{i\al \phi (z, \zb ) } ~ e^{i\al' \phi (0) } |0> . $$
Then
$$ \d_z \d_\zbar \Sigma = \frac{m^2}{2} (1- \cosh 2 \eta ) $$
$$ \d_z \d_\zbar \eta = \frac{m^2}{2}  \sinh 2\eta . $$

Note that the above differential equations are do not depend on
$\al , \al'$, though the correlation function itself certainly does.
This implies that the $\al, \al'$ dependence must come in as
a boundary condition for the solution of the differential equation.

An important question is to understand whether the above differential
equations can somehow be obtained from the affine $\hat{sl(2)}$
symmetry.  On the one hand, the above differential equations
are known to be part of the differential equations of the
$\hat{sl(2)}$ hierarchy.  However we know of no concrete connection
between the later affine structure and the genuine affine symmetry
of the quantum field theory.  It would be very interesting to
relate these two structures since this may provide some ideas
on how to $q$-deform the above results on correlation functions,
though this is rather speculative.

\newsec{Translation Anomaly and Short Distance Expansion}

The infinite integral representation (Fredholm determinant) that
we obtained in the last section
for the 2 point functions using the resolution of
the identity and the form factors is a large distance expansion.
The symmetry $\slhh$ does not easily relate various terms in this
 form factor sum since it commutes with particle number.
However it seems more likely that the $\slhh$ symmetry could relate
various terms in the short distance expansion.  One reason for
believing this is that the short distance expansion is just conformal
perturbation theory, and the quantum affine charges were originally
derived in this framework, however we have very little that is concrete
to add to this speculation.

We would like to show how one can use the infinite Heisenberg algebras
described above to recover the short distance expansion of a
simple correlator.   In the radial quantization, the integrals of motion
split into chiral pieces each satisfying infinite Heisenberg algebras:
\eqn\eVxxxvi{
\[ \al^L_n , \al^L_m \] = n~ \ma^{2|n|} ~ \de_{n,-m} , ~~~~~
\[ \al^R_n , \al^R_m \] = n~ \ma^{2|n|} ~ \de_{n,-m} . }
For the translation operators, this implies
$$P_z  = \al^L_{-1} + \al^R_1  ,  ~~~~~P_\zbar = \al_1^L + \al_{-1}^R .  $$
Of course the translation operators commute: $[P_z , P_\zbar ] = 0$,
however there are anomalies in the split pieces:
$[\al_1^L , \al_{-1}^L ] = [\al_1^R , \al_{-1}^R ] = \ma^2 $.
The idea is the use the non-manifest translation invariance in
radial quantization to our advantage.

We will consider the correlation function:
\eqn\ecorr{
\langle \ha | \psi_+ (z, \zbar) \psi_- (w , \bar{w} ) | - \ha \rangle
= \inv{z-w} \sqrt{ \frac{w}{z} } + \CO (\ma^2 ) +...... }
In conformal perturbation theory all of the higher order terms
in powers of $m$  have complicated integral representations.
Namely,
\eqn\shh{
<-\ha | \psi_+ (z, \zbar) \psi_- (w , \bar{w} ) |  \ha >
= \sum_{n=0}^\infty  \( \frac{\ma}{\pi} \)^{2n}
\inv{(2n)!}  \left( \matrix{2n\cr n\cr }\right)
\int d^2 z_1  \cdots d^2 z_{2n} ~ C(z,\zb , w, \bar{w} ;
z_1, \cdots z_{2n}  ) , }
where
\eqnn\ecc
$$\eqalignno{
C = \sqrt{\frac{z}{w} } \inv{z-w}
&
\frac{ \[ \prod_{i,j = 1,..n, i\neq j} |z_i - z_j |^2 \]
\[ \prod_{i,j = n+1...2n, i\neq j} |z_i - z_j|^2  \] }
{\[ \prod_{r=1}^n \prod_{s=n+1}^{2n} |z_r - z_s |^{-2} \] }
&\ecc\cr
&  \times \[  \prod_{i=1}^n
\frac{z_i - z}{z_i - w}
\frac{z_{i+n} - \om}{z_{i+n} - z }
\frac{ |z_i|}{|z_{i+n} | }  \]  . \cr }$$
We will give a simple operator construction of all of these
higher order terms.

One starts from the relation
\eqn\etrans{
\psi (z, \zbar ) = e^{z \al_{-1} + \zbar \al_1 }
{}~ \psi (0) ~
e^{-z \al_{-1} - \zbar \al_1 } . }
As we have seen, at $r=0$, we are very close to conformal field
theory.  Let us take the expression:
\eqn\ero{
\psi (0) =  \lim_{\ep , \bar{\ep} \to 0} ~
\sum_\omega  b_\om  \ep^{-\om - 1/2}
+ \frac{\ma}{\ha - \om }  \bb_\om \bar{\ep}^{\ha - \om }  . }
The first term above is just as in conformal field theory; the
second term can be obtained from first order perturbation theory.
Using
\eqn\baker{
e^A \psi e^{-A} = \psi + [A, \psi ] + \ha [A, [A, \psi ]] + ....}
and
$$ [\al_{-1}^L , b_\om ] = (\ha - \om ) b_{\om -1} , ~~~~~
[\al_1^L , b_\om ] = - \frac{\ma^2}{\ha + \om }  b_{\om + 1} $$
$$ [\al_{-1}^R , \bb_\om ] = (\ha - \om ) \bb_{\om -1} , ~~~~~
[\al_1^R , \bb_\om ] = - \frac{\ma^2}{\ha + \om }  \bb_{\om + 1} $$
one obtains
\eqn\epsi{
\psi( z, \zbar ) = \sum_{\om \in \Zmath}  b_\om f_\om (z, \zbar)
+ \bb_\om \bar{f}_\om (z, \zbar) .}
The functions $f_\om , \bar{f}_\om$  are easily computable order
by order in $\ma$ from \etrans\baker.  One obtains
\eqn\efom{
f_\om = z^{-\om - \ha} \( \sum_{n=0}^\infty (-1)^n
\frac{ (\ma^2 z\zb )^n }{n!}  \frac{ \Gamma (\ha + \om -n )}{\Gamma (\ha +
\om)}
\) }
\eqn\efomb{
\bar{f}_\om
= \ma ~ \zb^{\ha - \om} \sum_{n=0}^\infty (\ma^2 z\zb)^n \inv{n!}
\frac{ \Gamma (\ha - \om )}{\Gamma (\frac{3}{2} - \om + n )}  }
One recognizes these functions as short distance expansions of
the functions:
\eqn\fbes{
f_\om = \Gamma (1/2 - \om ) \ma^{\om + \ha}
\( \frac{\zb}{z} \)^{(\om + 1/2)/2} I_{-\ha - \om } (\ma r)   }
\eqn\fbesb{
\bar{f}_\om = \Gamma (1/2 - \om ) \ma^{\om + \ha}
\( \frac{\zb}{z} \)^{( 1/2-\om)/2} I_{\ha - \om } (\ma r)   }
Finally, inserting these expansions into the correlation function,
one obtains:
\eqn\eIIIxxxxiv{\eqalign{
\lva{\ha} ~ \psi_+ & (r, \vphi ) ~ \psi_- (r' , \vphi ' ) ~
\va{-\ha}_{\vphi = \vphi' = 0}   \cr
& =~ \ma \pi \[ \( \sum_{n=1}^\infty (-)^n ( I_{-n-\ha} (\ma r ) I_{n-\ha}
(\ma r' ) - I_{\ha-n} (\ma r) I_{\ha + n} ( \ma r' )) \)
- I_\ha (\ma r) I_\ha (\ma r' ) \] \cr
& =~ \mtoo ~ \sqrt{\frac{z}{z'}} \> \inv{z-z'} . \cr }}

To summarize what we think is interesting, we have shown how from
structures in radial quantization we can reconstruct the short distance
expansion of a correlation function in an operator framework.
It would be very interesting to generalize this to theories other
than free fermions, though this seems rather difficult.

Finally we remark that the above formulation of a correlation function
is reminiscent of the free fermion construction of tau functions\ref\rtau{E.
Date, M. Kashiwara, M. Jimbo and T. Miwa, {\it Transformation Groups
for Soliton Equations}, Proceedings of RIMS Symposium, Kyoto, Japan,
May 1981, M. Jimbo and T. Miwa, eds, World Scientific Publishing Co.,
Singapore, 1983.}.  In fact it is natural to incorporate the
dependence on higher coordinates as follows:
$$  \tau( z_1, \zb_1, z_2 , \zb_2 ,.....)
    = <0| \exp ( \sum_n \al_n z_n )  \Phi (0)  \exp (-\sum_n \al_n z_n )
    |\Phi' >   $$

The short distance expansion of general sine-Gordon correlation functions
is studied in \ref\rkon{R. Konik and A. LeClair, {\it Short Distance
Expansion of Sine-Gordon Correlation Functions}, in preparation.}

\newsec{Discussion}

Though we have limited ourselves to perhaps the simplest possible
case of the free-fermion point of the sine-Gordon theory, we
believe the ideas presented here can lead to a new framework for
computing form factors in massive integrable quantum field theory.
In this approach, since a complete description of the space of
fields $\hf$ is provided from the outset via radial quantization,
the complete set of solutions to the form factor bootstrap is
automatically yielded.
In the bosonized construction in section 5, an important role was
played by the affine $\slh$ symmetry.  Since away from the free-fermion
point this symmetry is deformed to a $\CU_q (\slh )$
symmetry\ref\rnlc{\BLnlc},
this quantum affine symmetry is expected to be important for
the general construction.  The results contained in
\ref\rfr{\FrResh}\ref\rdefkz{F. A. Smirnov, Int. J. Mod. Phys. A7,
Suppl. 1B (1992) 813.}\ref\rfj{\Frenkeli}\ should prove
useful.

It is interesting the compare the basic features of our construction
with Lukyanov's
approach \ref\rluka{S. Lukyanov, {\it Free Field Representation for Massive
Integrable Models}, Rutgers preprint RU-93-30, hep-th/9307196.}
\ref\rlukb{S. Lukyanov, Phys. Lett. B325 (1994) 409.},
(see also \ref\shat{S. Lukyanov and
S. Shatashvili, Phys. Lett. B298 (1993) 111.} ), where
the form factors are constructed as
traces over auxiliary Fock-modules.
The original motivation behind his construction came
from the work\ref\vvstar{\VVstar}, where the necessary
mathematical
properties of these traces were understood in the
context of lattice models.

Though the constructions above  and \rluka\ are similar in
spirit, the detailed aspects of the constructions are quite
different;  as mentioned above, here the form factors are
vacuum expectation values of vertex operators whereas in
\rluka\ they are traces  over infinite dimensional modules
of products of vertex operators.  Namely, in \rluka\ the n-particle
form factors of a field $\Phi$ are represented as
\eqn\trace{
Tr_\CF \( e^{2\pi i L} V (\th_1 ) \cdots V (\th_n ) \> \CO_{\Phi} \)  }
where $\CF $ is an infinite dimensional auxiliary Fock space,
$V(\th )$ are vertex operators depending on the rapidity $\th$,
$L$ is the generator of Euclidean rotations (Lorentz boost),
and $ \CO_{\Phi} $ is an operator that encodes the data of the
field $\Phi$.

A completely satisfactory understanding of the
origins of these differences is lacking; indeed, the
specialization of the results in \rluka\ to the free-fermion
point has not been carried out, and furthermore is not an
easy exercise.  Nevertheless, the main differences in the
constructions may be understood heuristically as arising
from the distinction between {\it radial} and {\it angular}
quantization.  Let us clarify this point.
Define the usual Euclidean light-cone and polar coordinates  as
follows
\eqn\coor{
z = (t+ix)/2 =\frac{r}{2} \exp(i\vphi) , ~~~~~~~~
\zbar = (t-ix)/2 = \frac{r}{2} \exp (-i\vphi ) . }
Our work was carried out
in radial quantization, where  $r$ is declared as the `time'.
In angular quantization, $\vphi$ is declared as the `time'.
By considering the free-boson limit of the
sine-Gordon theory, where an explicit angular quantization
may be performed, it was shown in \rlukb\ how the main features
of the algebraic contruction in \rluka\ arise.
In particular, the factor $\exp 2\pi i L$ in the trace was interpreted
as a density matrix.

The origin of the traces in angular quantization can also be
understood heuristically as follows.
In angular quantization, since the Lorentz boost operator $L$
generates shifts in $\vphi$, it is the Hamiltonian.
Consider now the functional integral formulation of such a quantization
scheme.  To do this, one must cut the spacetime plane with a semi-infinite
line from the origin to infinity.  Equal time contours are circles surrounding
the origin, where the initial and final times correspond to the
two sides of the above semi-infinite line.  From the standard
correspondence between path integrals and quantum matrix elements,
when one identifies the states on each side of the semi-infinite line
and sums over them, one obtains:
\eqn\func{
\lvac ~ \CO ~ \rvac  = \frac{\int ~ D\Phi  ~ e^{-S} ~ \CO}
{\int D\Phi e^{-S} }
{}~ = ~  \frac{Tr ~ ( e^{2\pi i L} \> \CO )}{Tr ~ ( e^{2\pi i L} )} . }
The $2\pi i$ constant in the factor $\exp (2\pi i L )$
is fixed by the
$2\pi$ length of the `time' $\vphi$.  In \vvstar\rluka,
the latter constant was fixed by imposing  the right
symmetry properties of the form factors expressed as these
traces.

Readers with some familiarity with conformal field theory will
doubtless see the strong parallels of this subject with the
work presented here.  For the example we have developed,
we have shown that in radial quantization
form factors can be computed as correlation functions in momentum space,
and these correlation functions are very similar in structure to
conformal spacetime correlation functions.  Furthermore, for the
purposes of computing form factors, one can describe the space of fields
in the same way as is done in the ultraviolet conformal field theory.
In a definite sense, we have shown that starting from a description
of the space of fields in a conformal field theory and the basic
operators from which one constructs this space (in our case, we mean
the operators $b^\pm_\om , \bb^\pm_\om $), then one can reconstruct
a massive theory and its form factors by constructing the vertex operators.
It is important to understand if this is possible more generally.

As far as correlation functions are concerned, it is disappointing
that thus far we have been unable to develop some new methods to
derive their main properties from the (quantum) affine symmetry.
It would be very interesting to understand even how the differential
equations of section 6 are related to the affine $\hat{sl(2)}$ symmetry.

\bigskip

\bigskip

\centerline{\bf Acknowledgements}

\bigskip

I would like to thank the organizers  for all of their
efforts toward this very successful meeting.  I also thank
my collaborators on the work presented here, Denis Bernard
and Costas Efthimiou.   Finally  I am grateful to S.H. for
southern hospitality in LA.
This work is supported by an Alfred P. Sloan Foundation fellowship,
and the National Science Foundation in part through the National
Young Investigator program.

\listrefs
\bye